\documentclass[twocolumn]{aastex701}
\usepackage{amsmath}
\usepackage{color}
\usepackage{booktabs}

\begin{document}

\title{Inherent self-consistency of the electron fraction between neutrino-dominated accretion flows and their progenitors}

\correspondingauthor{Tong Liu}
\email{tongliu@xmu.edu.cn}

\author[orcid=0009-0003-9659-6493,sname=Cui,gname=Rui-Qi]{Rui-Qi Cui}
\affiliation{Department of Astronomy, Xiamen University, Xiamen, Fujian 361005, China; tongliu@xmu.edu.cn}
\email{yqqnm2022@163.com}  

\author[orcid=0000-0001-8678-6291,sname=Liu,gname=Tong]{Tong Liu}
\affiliation{Department of Astronomy, Xiamen University, Xiamen, Fujian 361005, China; tongliu@xmu.edu.cn}
\email{tongliu@xmu.edu.cn}

\begin{abstract}
Stellar-mass black holes (BHs) surrounded by neutrino-dominated accretion flows (NDAFs) are a leading central engine of gamma-ray bursts (GRBs). In this work, we investigate the electron fraction distribution in NDAFs with or without disk outflows for different accretion rates, BH spins, and outflow rates. As the results, for the cases of the massive disks at relatively low accretion rates, the outer boundary of the disks are predominantly advection-cooled, yielding electron fractions of \(Y_{\rm e} \sim 0.5\), as expected for massive collapsar progenitors. By contrast, in the cases of lower-mass disk at high accretion rates, neutrino cooling becomes highly efficient and mildly electron-degenerate disks emerge, characterized by \(Y_{\rm e} \lesssim 0.38\) at the outer boundary of the disk, even for the strong outflows, which is consistent with materials from compact object merger scenarios. Moreover, we find that these trends remain robust across different BH spins. Consequently, the self-consistent agreement between the electron fraction properties at the outer boundaries of NDAFs and those expected from GRB progenitors provides effectively support for NDAFs serving as the GRB central engines.
\end{abstract}


\section{Introduction}
 
Gamma-ray bursts (GRBs) are among the most luminous explosive phenomena in the Universe. They are commonly classified into short- and long-duration bursts (SGRBs and LGRBs; e.g., \citealt{1993ApJ...413L.101K}). In general, SGRBs are thought to originate from the compact object mergers \citep[e.g.,][]{1986ApJ...308L..43P,1989Natur.340..126E,1992ApJ...395L..83N,2003MNRAS.345.1077R}, whereas LGRBs are typically associated with the core collapse of massive stars \citep[e.g.,][]{1993ApJ...405..273W,1999ApJ...524..262M}. Regardless of the progenitor system, the resulting central engine is usually a stellar-mass black hole (BH) surrounded by a hyperaccretion disk \citep[e.g.,][]{1991AcA....41..257P,1992ApJ...395L..83N,1999ApJ...524..262M}. However, although GRBs have been investigated for more than half a century, our understanding of their central engines remains relatively limited, particularly regarding the properties of the accompanying accretion disks.

Neutrino-dominated accretion flows (NDAFs) are a leading model for hyperaccretion disks \citep[e.g.,][]{1999ApJ...518..356P}. They have been considered in the context of GRB central engines because the luminosity produced by neutrino annihilation can naturally reach the level required to power GRBs \citep[e.g.,][]{2017NewAR..79....1L}. NDAFs have successfully explained several observed GRB features, such as the black-body components and the extended emission structure in prompt GRB emission \citep[e.g.,][]{2016ApJ...821..132L,2024ApJ...975..225L}. In addition, NDAFs are geometrically thick especially for the outer regions where the dimensionless disk half-thickness $H/R \sim 0.7-0.8$ \citep[e.g.,][]{2007ApJ...661.1025L,2007ApJ...657..383C} compared to standard thin disks with $H/R \ll 1$ \citep[e.g.,][]{2010A&A...521A..15A}. Although efficient neutrino cooling can reduce the inner-disk thickness where $H/R \lesssim 0.1-0.4$, $\nu \bar{\nu}$ annihilation remains strongly anisotropic and is preferentially concentrated toward the polar region around the angular-momentum axis \citep[e.g.,][]{2007ApJ...661.1025L}. This geometry naturally leads to the formation of a structured jet. Furthermore, when disk precession is taken into account, a precessing jet can be produced, which in turn provides a natural explanation for several observational phenomena, including quasi-periodic variability and specific variability shapes \citep[e.g.,][]{1999AAS..138..503P,2010A&A...516A..16L,2014MNRAS.441.2375H,2014ApJ...781L..19H}, X-ray plateau phases \citep[e.g.,][]{2021ApJ...916...71H}, and the low polarization observed in early optical afterglows \citep[e.g.,][]{2022ApJ...933..103H}.

Previous studies have progressively established how the electron fraction $Y_{\rm e} \equiv n_{\rm p}/(n_{\rm p}+n_{\rm n})$ of NDAFs depends on accretion regimes and neutrino emissions. \citet{2007ApJ...657..383C} developed a relativistic structure and showed that the inner disk self-regulates to a mildly degenerate state with $Y_{\rm e}\sim0.1$. \citet{2013ApJS..207...23X} found $Y_{\rm e}\approx0.46$ in the outer disk ($\geq 200~R_g$, where $R_g=2GM/c^2$ is the Schwarzschild radius and $M$ is the BH mass) by using a one-dimensional radial global relativistic solution. \citet{2013ApJ...762..102L} demonstrated a gradient in the vertical direction from the equatorial plane to the surface. The midplane became more neutronized by Urca processes, whereas the surface approached near proton--neutron symmetry. \citet{2026ApJ...996..142H} reported that $Y_{\rm e}$ versus several critical accretion rates over a wide range of BH masses exhibited a low-$Y_{\rm e}$ valley near the neutrino-opacity threshold. \citet{2024ApJ...971L..34M} presented a timescale-competition prescription for the disk electron fraction, showing that the outer regions of collapsar disks retain high $Y_{\rm e}$ $\sim 0.5$ close to initial, when the weak-interaction neutronization timescale was longer than the local accretion time. \citet{2020PhRvD.102l3014F} showed that long timescale viscous and neutrino radiation with prescribed initial disk masses further indicated progenitor dependence. Merger-like disks with $\sim0.1\,M_\odot$ yielded the mean outflow $\langle Y_{\rm e}\rangle\simeq0.30$--$0.35$, whereas collapsar-like disks with $\sim3\,M_\odot$ yielded $\langle Y_{\rm e}\rangle\gtrsim0.4$. 

Existing NDAF studies have characterized the distribution of the electron fraction, but a dedicated correspondence between $Y_{\rm e}$ and progenitors has been largely missing. Such a link is important for assessing engine viability because the expected electron fraction from progenitor-based studies provides an external benchmark for comparison, while $Y_{\rm e}$ of disk outflows largely determines the nucleosynthetic outcome and thus enables testable predictions for electromagnetic counterparts. For example on progenitor-based studies, \citet{2015PhRvD..91l4021F} performed a full-general-relativistic simulation of the post-merger evolution of a BH-neutron star (NS) binary and found that polar ejecta produced during disk circularization is fairly neutron rich. \citet{1995ApJS..101..181W} computed a suite of massive-star evolution models and parametrized supernova explosions, and their presupernova structure indicated that the electron fraction in the outer shells remained very close to 0.5.

In this paper, by varying the mass accretion rates, the disk outflow rates, and BH spins, we calculate the radial distribution of the electron fraction. Although the outer regions of NDAFs serve as the primary launching sites for disk outflows, a robust link of the electron fraction between the outer boundary of NDAFs and their progenitors should be expected. Sections \ref{sec:Model} and \ref{sec:Results} present the detailed BH-NDAF model and the results on the structure and electron fraction of NDAFs with or without outflows, respectively. A brief summary is made in Section \ref{sec:Summary}.

\section{Model}\label{sec:Model}
\subsection{Hydrodynamics}

We take central Kerr BHs into account and $a_{*}=Jc/(GM^2)$ is the dimensionless BH spin parameter that is defined by the BH mass $M$ and the angular momentum $J$. Viscous stress drives accretion in the disk, and the kinematic viscosity coefficient is $\nu=\alpha c_{\rm s}H$ where $H$ is the half-thickness of the disk, $c_{\rm s}=(P/\rho)$ is the isothermal sound speed with total pressure $P$ and mass density $\rho$, and $\alpha$ is a dimensionless constant which contains all the microphysics of viscous processes. In terms of conservation of mass, the accretion rate is obtained as
\begin{align}
\dot{M}=-4\pi Rv_r\rho H.
\end{align}
    
For the accreting Kerr BH, the relativistic correction factors have been given by \citep[e.g.,][]{1995ApJ...450..508R}:
\begin{align}
A&=1-\frac{2GM}{c^2R}+\left( \frac{GMa_{*}}{c^2R} \right)^2, \\
B&=1-\frac{3GM}{c^2R}+2a_{*} \left(\frac{GM}{c^2R} \right)^{3/2}, \\
C&=1-4a_{*} \left(\frac{GM}{c^2R} \right)^{3/2}+3 \left(\frac{GMa_{*}}{c^2R} \right)^2, \\
D&=\int_{R_{\rm{ms}}}^{R}\frac{\frac{x^2c^4}{8G^2}-\frac{3xMc^2}{4G}+\sqrt{\frac{a_{*}^2M^3c^2x}{G}}-\frac{3a_{*}^2M^2}{8}}{\frac{\sqrt{Rx}}{4} \left( \frac{x^2c^4}{G^2}-\frac{3xMc^2}{G}+2\sqrt{\frac{a_{*}^2M^3c^2x}{G}} \right) }dx,
\end{align}
where $R_{\rm{ms}}$ is the inner boundary of the disk. We take $R_{\rm inner} \simeq R_{\rm ms}=(3+Z_2-\sqrt{(3-Z_1)(3+Z_1+2Z_2)})R_{\rm g}$, where $R_{\rm ms}$ is the marginally stable orbit radius, $Z_1=1+(1-a_{*}^2)^{1/3}[(1+a_{*})^{1/3}+(1-a_{*})^{1/3}]$, and $Z_2=\sqrt{3 a_{*}^2+Z_1^2}$ \citep[e.g.,][]{1972ApJ...178..347B,2008bhad.book.....K,2021ApJ...920....5L}. The equation of the conservation of mass remains valid, and the hydrostatic equilibrium in vertical direction corrects the half-thickness of the disk \citep[e.g.,][]{1995ApJ...450..508R}:
\begin{align}
H\simeq \sqrt{\frac{PR^3}{\rho GM}}\sqrt{\frac{B}{C}}.
\end{align}
Conservation of angular momentum in the direction perpendicular to the disk leads to 
\begin{align}
f_{\mathrm{\phi}} (2\pi R \cdot 2H) \cdot R=\frac{\Delta L}{\Delta t},
\end{align}
where $f_{\rm{\phi}}$ is the viscous stress, $L$ is the disk angular momentum component in the direction perpendicular to the disk midplane, and $\Delta L/\Delta t=\dot{M}\Omega R^2$, where $\Omega =\sqrt{GM/R^3}$ is the Keplerian angular velocity. The viscous shear $f_\phi$ is corrected by \citep[e.g.,][]{2006AA...454...11R}:
\begin{align}
f_{\rm{\phi}} =\alpha P\sqrt{\frac{A}{BC}}=\frac{\dot{M}}{4\pi H}\sqrt{\frac{GM}{R^3}}\frac{D}{A}.
\end{align}

\subsection{Thermodynamics}

Energy balance equation is represented by the equality of viscous heating and cooling rates
\begin{align}
Q_{\mathrm{vis}}=Q^-,
\end{align}
where $Q_{\mathrm{vis}}$ is defined as \citep[e.g.,][]{1995ApJ...450..508R}
\begin{align}
Q_{\mathrm{vis}}=\frac{3GM\dot{M}}{8\pi R^3}\frac{D}{B}, \label{eq10}
\end{align}
and $Q^-$ has four contributions:
\begin{align}
Q^-=Q_{\mathrm{photo}}+Q_{\mathrm{adv}}+Q_{\mathrm{\nu}}+Q_{\mathrm{rad}}.
\end{align}
In the case of high temperatures and densities, the radiative cooling $Q_{\mathrm{rad}}$ is negligible as compared with other contributions. The photodissociation of $\alpha$-particles constitutes a cooling mechanism, with the rate given by 
\begin{align}
Q_{\mathrm{photo}}=6.8\times 10^{28}\rho _{10}v_rH\frac{d X_{\mathrm{nuc}}}{d R} (\rm{cgs\ units}), \label{eq12}
\end{align} 
where $\rho_{10}\equiv \rho /(10^{10}\ \mathrm{g\ cm^{-3}})$ and $X_{\mathrm{nuc}}$ is the mass fraction of the free nucleons \citep[e.g.,][]{2005ApJ...629..341K}. The advective cooling rate is 
\begin{equation}
\begin{aligned}
Q_{\mathrm{adv}} &= \rho v_r H T \frac{ds}{dR} \\
&\approx -v_r \frac{H}{R} T \left( \frac{4}{3} a T^3 + \frac{3}{2} \frac{k_B \rho}{m_u}\frac{1+3X_{\mathrm{nuc}}}{4} + \frac{4}{3}\frac{u_{\nu}}{T} \right),
\end{aligned}
\end{equation}
where $s$ is the specific entropy, $T$ is the disk temperature, $a=4\sigma /c$ is the radiation constant and $\sigma$ is the Stefan-Boltzmann constant, and $m_u$ is the mean mass of a nucleon. The three terms in parentheses are the entropy density of photons, of free nucleons and $\alpha$-particles, and of neutrinos, respectively, and $u_{\nu}$ is the neutrino energy density \citep[e.g.,][]{2010A&A...516A..16L}:
\begin{align}
u_{\nu}=\sum_{i}\frac{\frac{7}{8}aT^4(\tau_{\nu _i}/2+1/\sqrt{3})}{\tau_{\nu _i}/2+1/\sqrt{3}+1/3\tau_{\rm a,\nu _i}}. \label{eq14}
\end{align}
The cooling rate due to neutrino loss can be expressed as \citep[e.g.,][]{2005ApJ...629..341K}:
\begin{align}
Q_{\mathrm{\nu}} =\sum_{i}\frac{\frac{7}{8}\sigma T^4}{\frac{3}{4}\left[ \tau_{\nu _i}/2+1/\sqrt{3}+1/(3\tau_{\mathrm{a},\nu _i}) \right]}. \label{eq15}
\end{align}
In Equations (\ref{eq14}) and (\ref{eq15}), $\tau_{\nu _i}$ is the total optical depth of neutrinos 
\begin{align}
\tau_{\nu _i}=\tau_{\mathrm{a},\nu _i}+\tau_{\mathrm{s},\nu _i}, \label{eq16}
\end{align}
where $\tau_{\mathrm{a},\nu _i}$ is the absorption optical depth of neutrinos, and $\tau_{\mathrm{s},\nu _i}$ is the optical depth of neutrinos through scattering off free nucleons, electrons, and $\alpha$-particles, where the subscript $i$ is the index of the neutrino species which include $\nu_{\mathrm {e}},~\nu_{\mathrm{\tau}}\ \rm{and}\ \nu_{\mathrm{\mu}}$. 
$\tau_{\mathrm{s},\nu _i}$ is given by
\begin{equation}
\begin{aligned}
\tau_{\mathrm{s},\nu _i}&=H/\lambda _{\nu _i}\\&=H[\sigma _{\rm p,\nu  _i}\tilde{n}_{\rm p}+\sigma _{\rm n,\nu  _i}\tilde{n}_{\rm n}+\sigma _{\alpha ,\nu  _i}n_{\alpha}+\sigma _{\rm e,\nu  _i}(n_{\rm e^-}+n_{\rm e^+})]. \label{eq17}
\end{aligned}
\end{equation}
In Equation (\ref{eq17}), $\lambda _{\nu _i}$ is the mean free path, $\sigma _{\rm p,\nu  _i}, ~\sigma _{\rm n,\nu  _i}, ~\sigma _{\alpha ,\nu  _i}$, and $\sigma _{\rm e,\nu  _i}$ are cross sections of scattering on protons, neutrons, $\alpha$-particles, and electrons, respectively, and $\tilde{n}_{\rm p}, ~\tilde{n}_{\rm n}, ~ n_{\alpha}, ~n_{\rm e^-}$, and $n_{\rm e^+}$ are the number densities of free protons, free neutrons, $\alpha$-particles, electrons, and positrons, respectively. The four scattering cross-sections are as follows \citep[e.g.,][]{2004ASSL..302..133B,2007ApJ...661.1025L}:
\begin{align}
\sigma _{\rm p,\nu  _i}&=\frac{1}{4}\sigma _0 E_{\nu  _i}^2[(C_{\rm V,\nu _i}-1)^2+3g_{\rm A}^2(C_{\rm A,\nu _i}-1)^2],\\
\sigma _{\rm  n,\nu  _i}&=\frac{\sigma _0 E_{\nu  _i}^2}{4} \frac{1+3g_{\rm A}^2}{4},\\
\sigma _{\rm  \alpha ,\nu  _i}&=4\sigma _0 E_{\nu  _i}^2 \rm {sin}^4\theta _{{\rm W}}, 
\end{align}
\begin{equation}
\begin{aligned}    
\sigma _{\rm e,\nu  _i}&=\frac{3 k_{\rm B} T \sigma_0 E_{\nu  _i}}{8m_{\rm e}c^2}\left(1+\frac{\eta _{\rm e}}{4}  \right)\\ &\times \left[ (C_{\rm V,\nu _i}+C_{\rm A,\nu _i})^2+\frac{1}{3}(C_{\rm V,\nu _i}-C_{\rm A,\nu _i})^2 \right], 
\end{aligned} 
\end{equation}
where $\sigma _0=1.76\times 10^{44} \ \rm{cm^2}$, $E_{\nu  _i}$ is the mean energy of neutrinos in units of $m_{\rm e} c^2$, and $g_{\rm A} \approx 1.26$, ~$C_{\rm V,\nu _e}=\frac{1}{2}+2\ \rm {sin}^2\theta _{{\rm W}}$, ~$C_{\rm V,\nu _{\mu}}=C_{\rm V,\nu _{\tau}}=-\frac{1}{2}+2\ \rm {sin}^2\theta _{{\rm W}}$, $C_{\rm A,\nu _{\rm e}}=C_{\rm A,\bar{\nu} _{\rm {\mu}}}=C_{\rm A,\bar{\nu} _{\rm {\tau}}}=\frac{1}{2}$, ~$C_{\rm A,\bar{\nu} _{\rm e}}=C_{\rm A,\nu _{\rm {\mu}}}=C_{\rm A,\nu _{\rm {\tau}}}=-\frac{1}{2}$, and $\rm {sin}^2\theta _{{\rm W}}\approx 0.23$. Fermi-Dirac integration provides the number density of negative and positive electrons \citep[e.g.,][]{2005ApJ...629..341K,2007ApJ...661.1025L}
\begin{align}
    n_{\rm e^{\mp}}=\frac{1}{\hbar ^3 \pi^2}\int_0^\infty\  \mathrm{d} p \ p^2\frac{1}{e^{(\sqrt{p^2c^2+m_{\rm e}^2c^4}\mp \mu _{\rm e})/k_{\rm B}T}+1}.
\end{align}

In Equation (\ref{eq16}), the absorption depth for neutrinos is expressed as
\begin{align}
    \tau_{\rm a,\nu _i}=\frac{q_{\nu _i}H}{4(7/8)\sigma T^4},
\end{align}
where $q_{\nu _i}$ is the total neutrino cooling rate and concludes four terms,
\begin{align}
    q_{\nu _i}=q_{\rm Urca}+q_{\rm e^{\rm -}e^{\rm +}}+q_{\rm brem}+q_{\rm plasmon}.
\end{align}
In the cooling rate, the Urca process is only related to neutrinos and consists of these three items \citep[e.g.,][]{2005PhRvD..72a3007Y,2007ApJ...661.1025L},  
\begin{align}
    q_{\rm Urca}=q_{\rm p+e^-\rightarrow n+\nu _{\rm e}}+q_{\rm n+e^+\rightarrow p+\bar{\nu }_{\rm e}}+q_{\rm n\rightarrow p+e^-+\bar{\nu }_{\rm e}},
\end{align}
with 
\begin{equation}
\begin{aligned}
   q_{\mathrm{p+e^-\rightarrow n+\nu _{e}}}=&\frac{G_{\mathrm{F}}^2 \cos^2\theta_{\mathrm{c}}}{2\pi^2\hbar^3c^2} \left(1+3g_{\mathrm{A}}^2\right) \tilde{n}_{\mathrm{p}}\\
   &\!\!\!\!\!\times \int_Q^\infty \mathrm{d}E_\mathrm{e}E_\mathrm{e}\sqrt{E_\mathrm{e}^2-m_e^2c^4} \left(E_\mathrm{e}-Q\right)  ^3f_{\rm e^-},
\end{aligned}
\end{equation}
\begin{equation}
\begin{aligned}
   q_{\rm n+e^+\rightarrow p+\bar{\nu} _{\rm e}}=&\frac{G_{\mathrm{F}}^2 \cos^2\theta_{\mathrm{c}}}{2\pi^2\hbar^3c^2}\left(1+3g_{\mathrm{A}}^2\right) \tilde{n}_{\mathrm{n}}\\
   &\!\!\!\!\!\times \int_{m_{\mathrm {e}}c^2}^\infty \mathrm{d}E_\mathrm{e}E_\mathrm {e}\sqrt{E_\mathrm{e}^2-m_e^2c^4} \left(E_\mathrm{e}+Q\right)  ^3f_{\rm e^+},
\end{aligned}
\end{equation}
\begin{equation}
\begin{aligned}
   q_{\rm n\rightarrow p+e^-+\bar{\nu }_{\rm e}}\\
   &\!\!\!\!\!\!\!\!\!\!\!=\frac{G_{\mathrm{F}}^2 \cos^2\theta_{\mathrm{c}}}{2\pi^2\hbar^3c^2}\left(1+3g_{\mathrm{A}}^2\right)\\
   &\!\!\!\!\!\!\!\!\!\!\!\times \int_{m_{\mathrm {e}}c^2}^Q\mathrm{d}E_\mathrm{e}E_\mathrm {e}\sqrt{E_\mathrm{e}^2-m_e^2c^4} \left(Q-E_\mathrm{e}\right)^3(1-f_{\rm e^-}),
\end{aligned}
\end{equation}
where $G_{\rm F}\approx 1.436\times 10^{-49}\ \mathrm{erg\ cm^3}$, $\mathrm{cos}^2\theta _\mathrm{c}\approx 0.947$, $Q=(m_\mathrm{n}-m_\mathrm{p})c^2$, and the Fermi-Dirac function is $f_{\mathrm{e}^{\mp}}=\{\mathrm{exp}[(E_\mathrm{e}\mp \mu _{\rm e})/k_{\rm B}T]+1\}^{-1}$. The electron-positron pair annihilation rate into neutrinos can be expressed as \citep[e.g.,][]{2005ApJ...629..341K}
\begin{align}
    q_{\rm e^{\rm -}e^{\rm +}\rightarrow \nu_{\rm e}\bar{\nu }_{\rm e}}\approx 3.4\times 10^{33}T_{11}^9\ \mathrm{erg\ cm^{-3}\ s^{-1}},
\end{align}
\begin{equation}
\begin{aligned}
q_{\rm e^{\rm -}e^{\rm +}\rightarrow \nu_{\rm \mu}\bar{\nu }_{\rm \mu}}&=q_{\rm e^{\rm -}e^{\rm +}\rightarrow \nu_{\rm \tau}\bar{\nu }_{\rm \tau}} \\ &\approx 0.7\times 10^{33}T_{11}^9\ \mathrm{erg\ cm^{-3}\ s^{-1}}.
\end{aligned}
\end{equation}
Through the process $n+n\rightarrow n+n+\nu _i+\bar{\nu }_i$, the nucleon-nucleon bremsstrahlung rate, $q_{\rm brem}$, is the same for three species of neutrinos \citep[e.g.,][]{2000ApJ...539..865B,1998ApJ...507..339H,2007ApJ...661.1025L},
\begin{align}
    q_{\rm brem}\approx 1.5\times 10^{27}\rho_{10}^2T_{11}^{5.5}\ \rm{erg\ cm^3\ s^{-1}}.
\end{align}
Finally, the plasmon decay rate $q_{\rm plasmon}$, only considers the process $\tilde{\gamma}\rightarrow \nu_{\rm e}+\bar{\nu }_{\rm e}$, where plasmons $\tilde{\gamma }$ are photons that interact with electrons \citep[e.g.,][]{1996AA...311..532R},
\begin{equation}
\begin{aligned}
    q_{\rm plasmon}&\approx 1.5\times 10^{32}T_{11}^9\gamma_{\rm p}^6\mathrm{exp}(-\gamma _{\rm p})\\
    &\times (1+\gamma_{\rm p}) (2+\frac{\gamma _{\rm p}^2}{1+\gamma _{\rm p}}),
\end{aligned}
\end{equation}
where $\gamma_{\rm p}=5.565\times 10^{-2}[(\pi ^2+3\eta_{\rm e}^2)/3]^{1/2}$.

The equation of state comprises four pressure components:
\begin{align}
P=P_{\rm{gas}}+P_{\rm{rad}}+P_{\rm{e}}+P_{\rm{\nu}}.
\end{align} 
The gas pressure from free nucleons and $\alpha$-particles is 
\begin{align}
P_{\rm{gas}}=\frac{k_{\rm{B}}\rho T}{m_u}\frac{1+3X_{\rm{nuc}}}{4}.\label{eq34}
\end{align} 
The photon radiation pressure is 
\begin{align}
P_{\rm{rad}}=\frac{aT^4}{3}.
\end{align}
The electron pressure comes from the sum of the contributions of positrons and electrons and we calculate it with the exact Fermi-Dirac distribution. It can be expressed as
\begin{align}
P_{\rm{e}}=P_{\rm{e^-}}+P_{\rm{e^+}}
\end{align}
with
\begin{equation}
\begin{aligned}
P_{\rm{e^{\mp}}}=&\frac{1}{3\pi ^2\hbar ^3c^3}\\ &\!\!\!\!\!\times \int_{0}^{\infty}\mathrm {d}p\frac{p^4}{\sqrt{p^2c^2+m_{\mathrm{e}}^2 c^4}}\frac{1}{e^{( \sqrt{p^2c^2+m_{\mathrm{e}}^2c^4}\mp \mu _{\mathrm {e}})/k_\mathrm {B} T}+1},
\end{aligned}
\end{equation}
where $\mu _{\mathrm{e}}$ is the chemical potential of electrons and can be measured by the degeneracy parameter and the electron degeneracy with the relationship $\mu _{\mathrm{e}}=\eta _{\mathrm{e}}k_{\mathrm{B}}T $. The neutrino pressure is 
\begin{align}
P_{\rm {\nu}}=u_{\rm {\nu}}/3.
\end{align}

\subsection{Electron Fractions}

\citet{2003ApJ...588..931B} demonstrated that $\beta$-equilibrium among free neutrons, free protons, and electrons in a hyperaccretion disk are established only if
$\dot{M}\gtrsim 10^{31}(\alpha /0.1)^{9/5}[M/(1 \ M_{\odot})]^{6/5}\ \rm{g\ s^{-1}}$.
The equilibrium composition depends sensitively on the neutrino optical depth.  In regions optically thick to neutrinos, trapped neutrinos enforce the reversible weak interactions,
\begin{align}
e^- + p &\rightleftharpoons n + \nu_{\rm e},\\
e^+ + n &\rightleftharpoons p + \bar\nu_{\rm e}, \label{eq40}
\end{align}
maintaining $\beta$-equilibrium. Conversely, in neutrino-transparent regions the inverse capture reactions are suppressed because neutrinos escape freely, so the system cannot sustain equilibrium in both directions. However,  the rate of the reaction $e^-+p\rightarrow n+\nu_{\rm e}$ is equal to the reverse reaction $e^++n\rightarrow p+\bar{\nu}_{\rm e}$, and the disk still attains $\beta$-equilibrium. To describe a transition regime, \citet{2007ApJ...661.1025L} introduced a weighting factor and derived an analytic relation between the electron fraction, \(Y_{\rm e}\), and the free-nucleon mass fraction, \(X_{\rm nuc}\):
\begin{align}
Y_{\rm e}=\frac{1-X_{\rm{nuc}}}{2}+\frac{X_{\rm{nuc}}}{1+\mathrm{exp}(\{[1+f(\tau_{\nu})]\mu _{\rm e}-Q\}/k_{\rm B}T)}.
\end{align}

\begin{figure}[!t]
\centering
\includegraphics[width=\columnwidth]{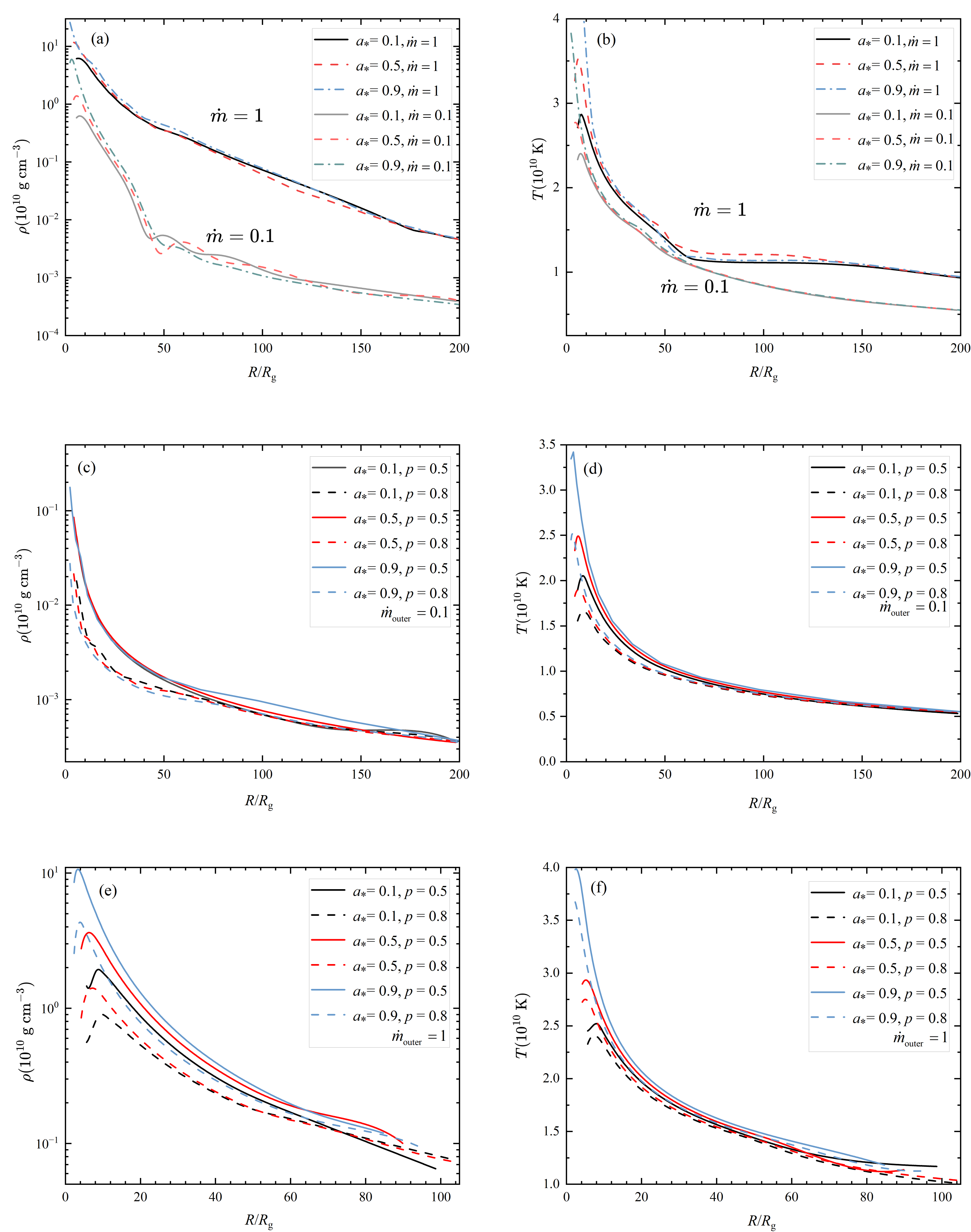}
\caption{Density and temperature profiles in (a)-(b) without outflows for $\dot{m}=0.1$ and $1$, and in (c)-(f) with outflows for $\dot{m}_{\rm outer}=0.1$ and $1$. Solid, dashed, and dot-dashed lines in (a) and (b) correspond to $a_*$ = 0.1, 0.5, and 0.9, respectively, while solid and dashed lines in (c)-(f) denote $p$ = 0.5 and 0.8, respectively.}
\label{fig1}
\end{figure}

The electronic chemical potential is constrained by the charge conservation law, and its relationship with the electron abundance is as follows:
\begin{align}
n_{\rm p}=\frac{\rho Y_{\rm e}}{m_{\rm u}}=n_{\rm e^-}-n_{\rm e^+}.
\end{align}

\begin{figure}[!t]
\centering
\includegraphics[width=\columnwidth]{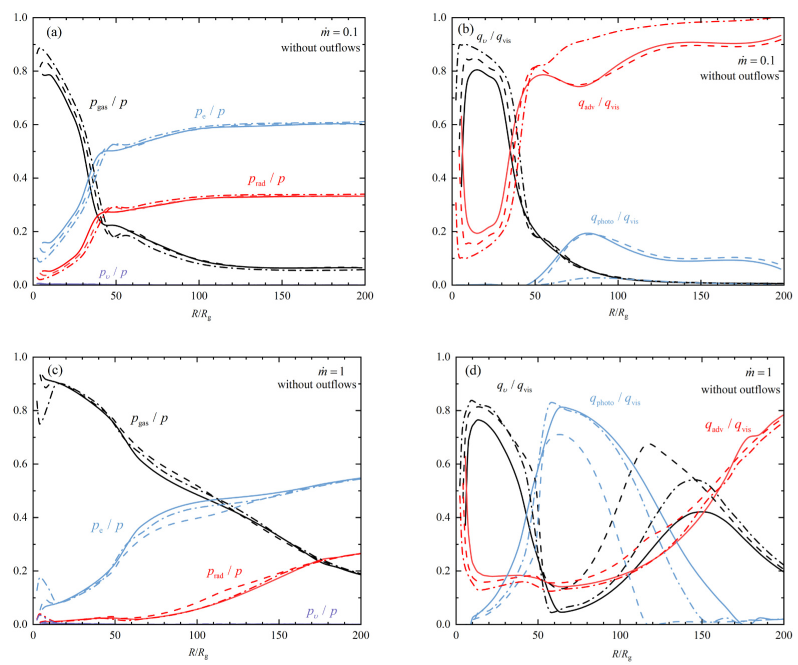}
\caption{Profiles of pressure contributions and cooling fractions of NDAFs without outflows with $\dot{m}$ = 0.1 and 1. Solid, dashed, and dot-dashed lines correspond to $a_*$ = 0.1, 0.5, and 0.9, respectively.}
\label{fig2}
\end{figure}

\section{Results}  \label{sec:Results}

We present numerical results for NDAFs models spanning a range of accretion rates, BH spins, and disk outflow strength representative of GRBs of both short and long duration. The dimensionless BH mass and accretion rate are defined by $m=M/M_{\odot}$ and $\dot m = \dot M/(M_{\odot}\ \mathrm{s}^{-1})$, respectively. The BH mass is fixed at $m=3$ and the viscosity coefficient at $\alpha=0.1$, while we vary the mass accretion rates $\dot{m}$ between 0.1, 1, and 5, and the dimensionless spin parameter $a_*$ between 0.1, 0.5 and 0.9. 

We also implement a parametric disk outflow which distributed with radius as a power-law \citep[e.g.,][]{1999MNRAS.303L...1B,2012ApJ...761..129Y,2014ARAA..52..529Y,2015MNRAS.453.3213S,2019MNRAS.482.2788S,2021ApJ...920....5L} 
\begin{equation}
\dot m=\dot m_{\rm outer}(R/R_{\rm outer})^{p},
\end{equation}
with $p=0.5$ or $0.8$, where $\dot{m}_{\rm outer}$ is the accretion rate at the disk's outer boundary $R_{\rm outer}$. In this section, we present the distribution of physical properties of the NDAFs as functions of radius. While for NDAFs without outflows, the disk mass inside of $R$ can be expressed by
\begin{equation}
M_{\rm disk} (R) = 4 \pi \int_{R_{\rm inner}}^{R} \rho H R dR. \label{eq44}
\end{equation} 
We focus on the electron fraction $Y_{\rm e}$ outcomes for various combinations of $\dot{M}$, $a_*$, and $p$, with particular attention to trends of electron fraction with accreted mass and radius.
 
\begin{figure}
\centering
\includegraphics[width=\columnwidth]{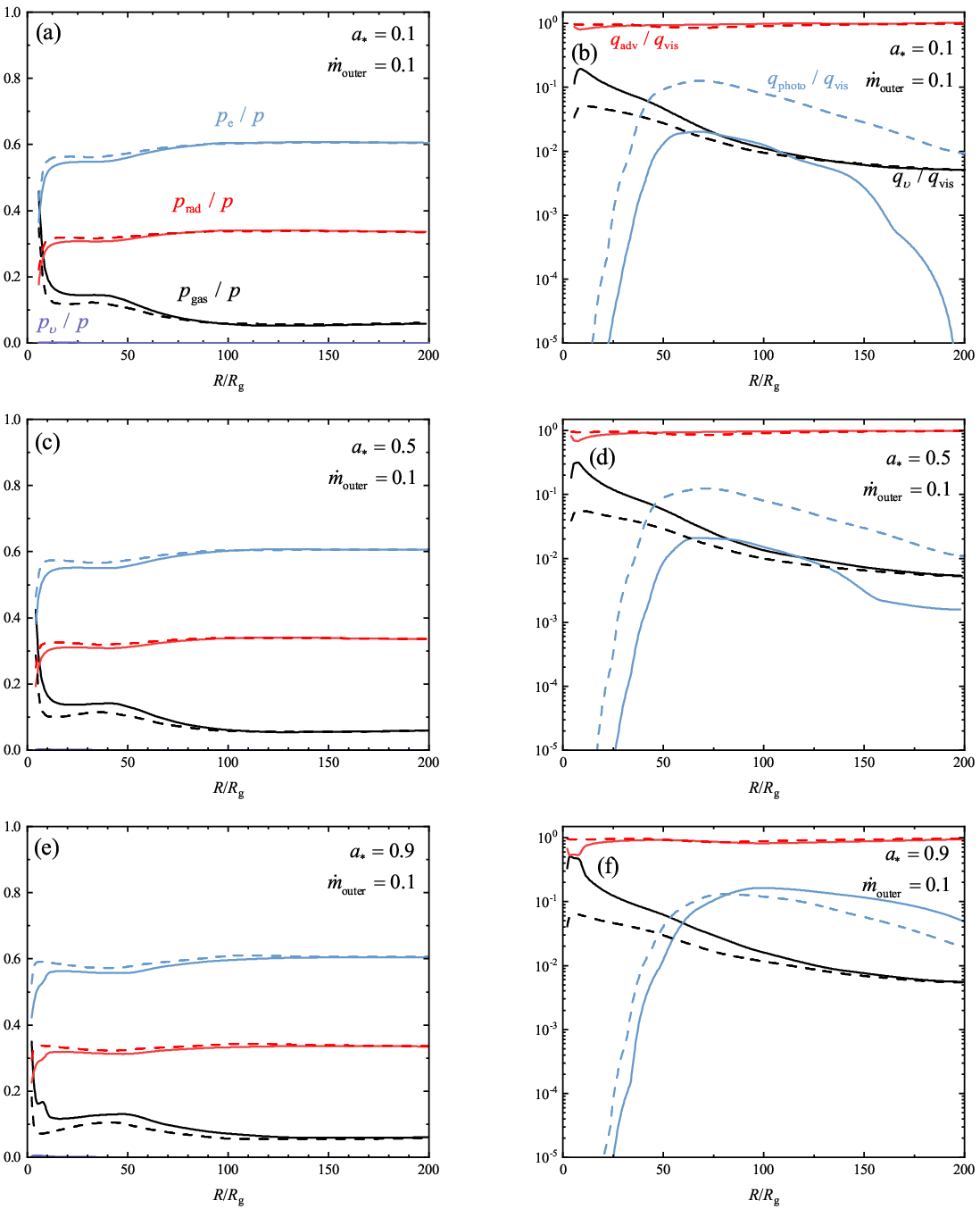}
\caption{Profiles of pressure contributions and cooling fractions with outflows for $\dot{m}_{\rm outer}=0.1$. Solid and dashed lines denote $p$ =  0.5 and 0.8, respectively.}
\label{fig3}
\end{figure}

\begin{figure}
\centering
\includegraphics[width=\columnwidth]{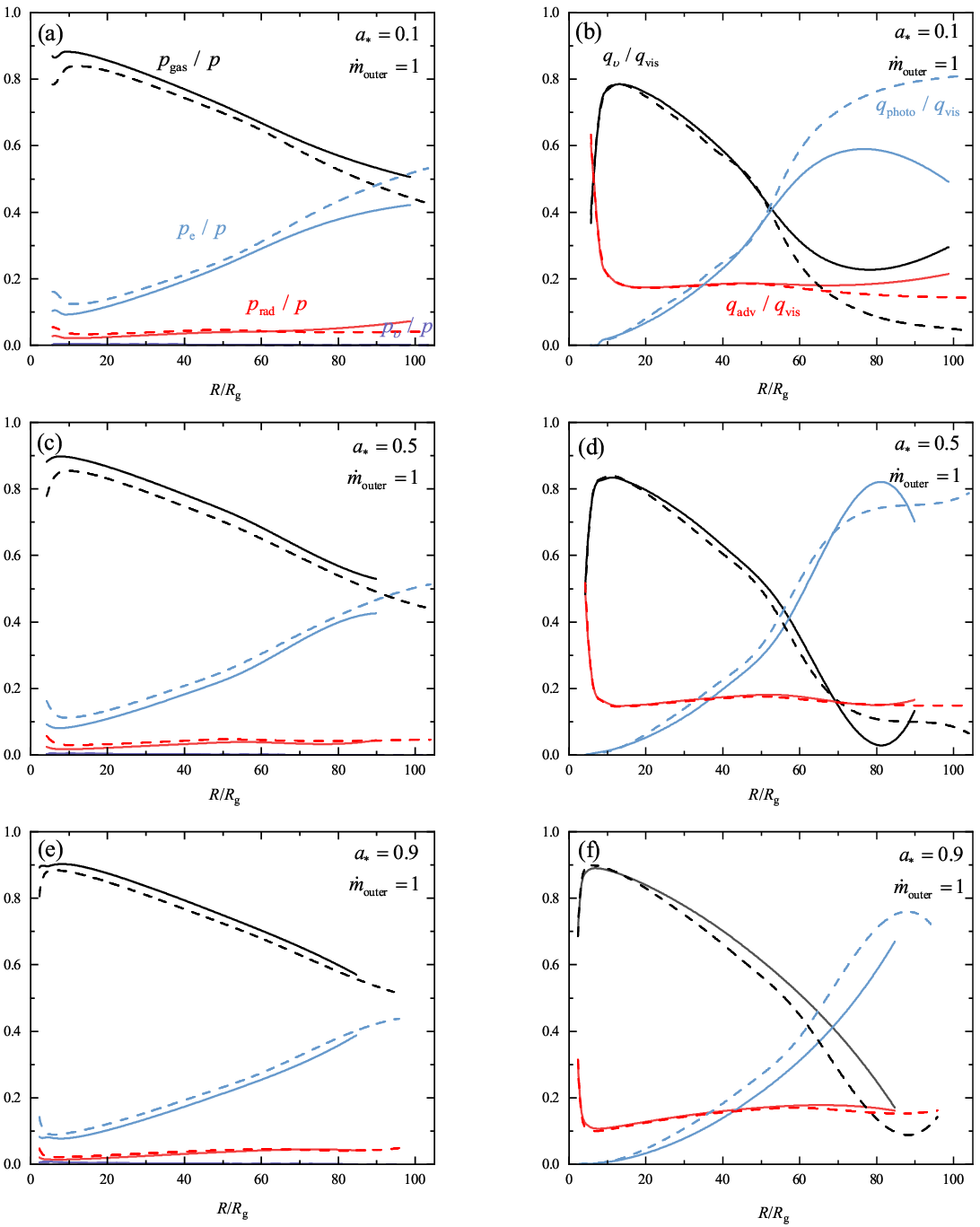}
\caption{Same as Figure \ref{fig3} except for $\dot{m}_{\rm outer}=1$.}
\label{fig4}
\end{figure}

\subsection{Structure}

In Figure \ref{fig1},  we show midplane radial profiles of density and temperature for $a_{\ast}=0.1, 0.5$, and $0.9$. Both $\rho$ and $T$ decrease monotonically with radius. Higher accretion rates and higher spins yield hotter and denser inner disks. Compared with Panels (a) and (b), Figures \ref{fig1}(c)-(f) show that outflows reduce the overall density and temperature but do not alter the high-temperature, high-density characteristics of inner regions. These trends are consistent with relativistic global NDAF solutions \citep[e.g.,][]{2013ApJS..207...23X}.

We then present radial profiles of pressure contributions and cooling fractions in Figure \ref{fig2} without outflows, and Figures \ref{fig3} and \ref{fig4} with outflows. For the pressure terms, we show the contributions of gas, photon radiation, electrons, and neutrinos. The contributions from neutrino pressure can be negligible for all cases. In Figures \ref{fig2}(a) and (c), at inner region of the disk, the temperature is of order $10^{10}~\mathrm{K}$, so helium nuclei are readily photodisintegrated, increasing the free-nucleon mass fraction $X_{\rm nuc}$. Pressure in NDAFs is dominated by baryons, and photon radiation pressure is always subdominant throughout although the photons are trapped within the disks. Only for very extreme accretion rates, the radiation pressure might be dominant as shown in Figure 6(c) of \citet{2005ApJ...629..341K}. At the larger radius, although the electron degeneracy weakens, the electron number density increase making $P_{\rm e}$ important toward the outer disk.

\begin{figure}[!t]
 \centering
\includegraphics[width=0.95\columnwidth]{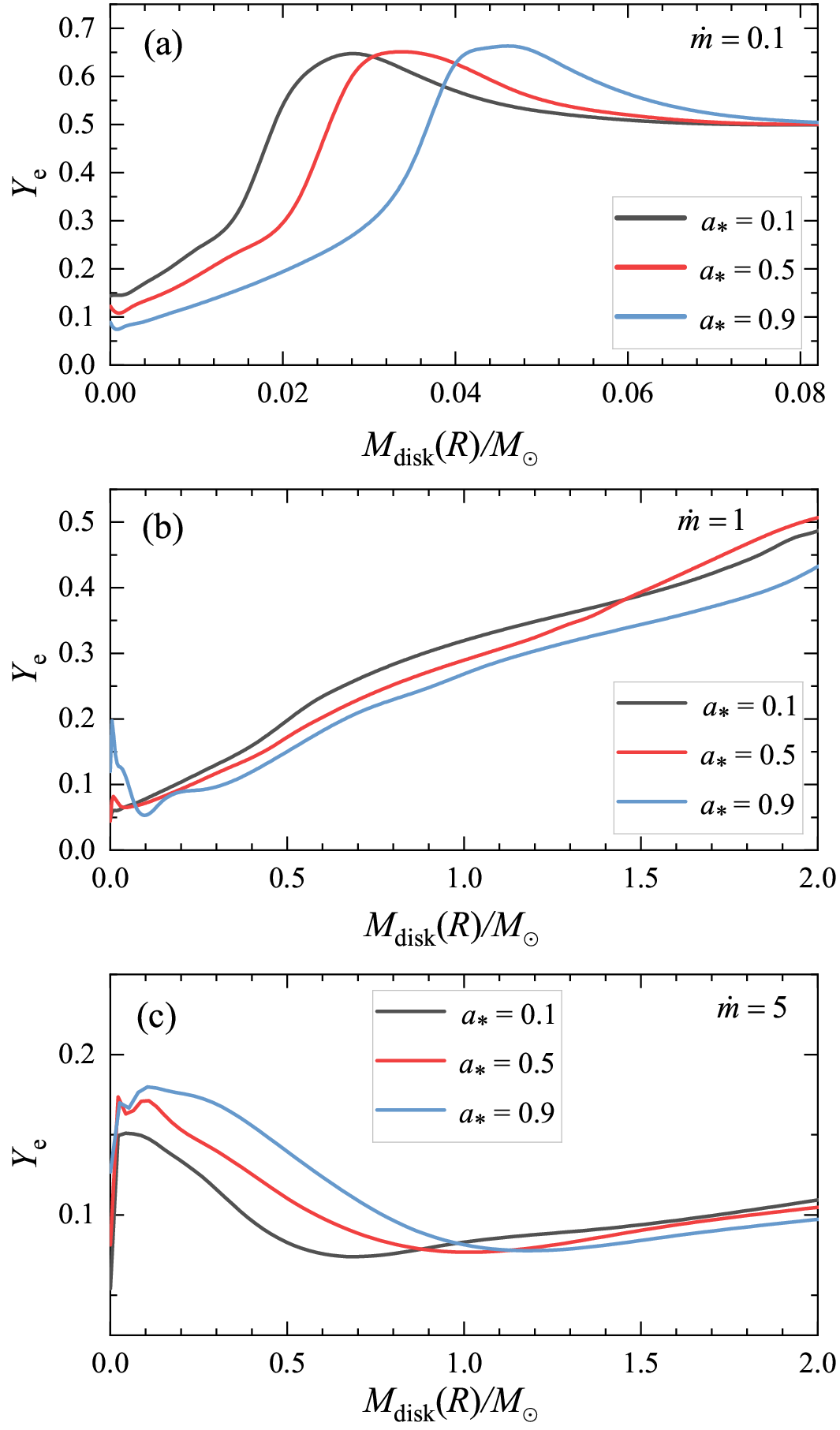}
\caption{Electron fraction $Y_{\rm e}$ of NDAFs without outflows as a function of disk mass $M_{\rm disk}(R)$ inside of $R$ for three accretion rates $\dot m$ = 0.1, 1, and 5, and black, red, and blue lines denote BH spins $a_\ast$ = 0.1, 0.5, and 0.9, respectively.}
\label{fig5}
\end{figure}

\begin{figure}[!t]
\centering
\includegraphics[width=0.95\columnwidth]{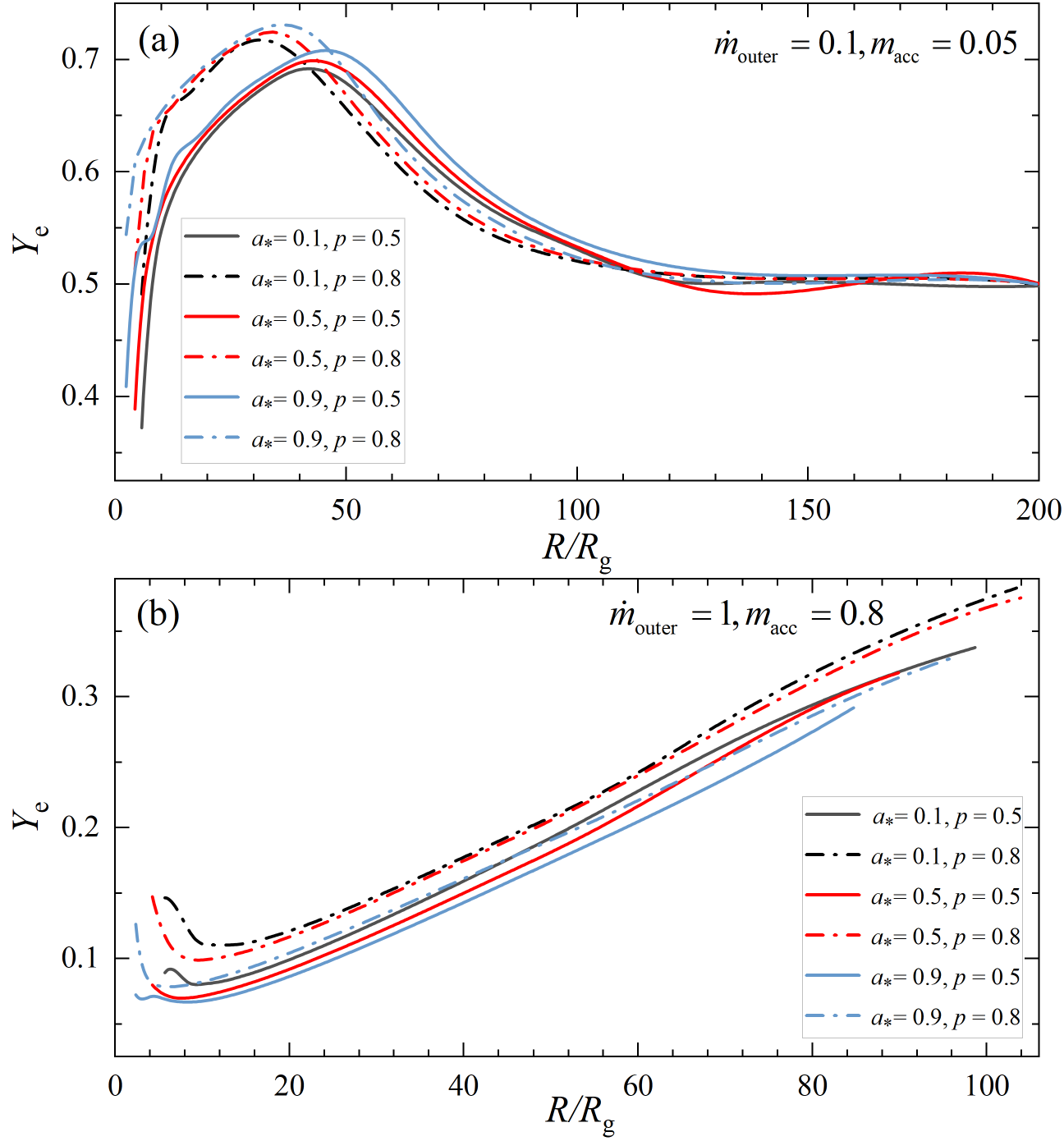}
\caption{Radial electron fraction $Y_{\rm e}$ of panel (a) and (b) for $\dot{m}_{\rm outer}$ = 0.1 and 1, and cumulative accreted mass $m_{\rm{acc}}=$0.05 and 0.8 $M_\odot$, respectively. The black, red, and blue lines denote BH spins $a_\ast$ = 0.1, 0.5, and 0.9, and solid and dashed lines correspond to outflow strength $p$ = 0.5 and 0.8, respectively.}
\label{fig6}
\end{figure}

For the energy terms, we plot the cooling rates of photodissociation, advection, and neutrino loss normalized to the viscous heating rate. At the low accretion rate, in Figure \ref{fig2}(b), the outer disks are dominated by advection, and in Figures \ref{fig3}(b), (d) and (f), the advection fractions increase further due to the presence of the disk outflows. At different accretion rates as shown in Figures \ref{fig2}(b) and (d), the cooling terms peak at different radii. Neutrino cooling rapidly becomes the dominant cooling channel in NDAFs near the ignition radius, $R_{\rm ign}$, where neutrino emission turns on and becomes important \citep[e.g.,][]{2007ApJ...657..383C,2017NewAR..79....1L}. Because the neutrino emissivity depends steeply on the local density and temperature, the neutrino-cooling efficiency decreases sharply outside $R_{\rm ign}$, and advective cooling correspondingly becomes the primary cooling mechanism \citep[e.g.,][]{1999ApJ...518..356P}. Photodisintegration cooling peaks around the $\alpha$-disintegration radius, $R_{\alpha}$, where $\alpha$ particles are dissociated into free nucleons (neutrons and protons), causing the $X_{\rm nuc}$ to vary rapidly with radius (i.e., Equation (\ref{eq12})). In Figure \ref{fig2}(b), $Q_{\rm photo}$ contributes a smaller fraction than in Panel (d), because disk at $\dot{m}=0.1$ is cooler and less dense, so $\alpha$-dissociation is weaker and confined to a narrower radial zone. The neutrino cooling rates peak in the inner disk and decline outward. In Figure \ref{fig2}(d), $Q_{\nu}$ exhibits secondary peaks in the outer disks. This occurs because neutrino cooling in the outer region remains dominated by the Urca process, whereas the photodisintegration term $Q_{\rm photo}$ declines rapidly with radii. At the same time, the viscous heating rates decrease steeply with increasing radii (i.e., Equation (\ref{eq10})). The combined reduction of $Q_{\rm photo}$ and $Q_{\rm vis}$ therefore makes the relative contribution of $Q_{\nu}$ rise again toward the disk edges.

\subsection{Electron fractions}

Figure \ref{fig5} shows the radial distribution of the local electron fraction as a function of the cumulative disk mass (i.e., Equation (\ref{eq44})) for NDAFs without outflows. In Figure \ref{fig5}(a), at $\dot m=0.1$ the electron fraction $Y_{\rm e}$ approaches 0.5 as the $M_{\rm disk}$ increases which indicates that protons and neutrons are nearly symmetric in the outer disk. While the typical accreted mass for long GRB progenitors is on the order of $1~M_\odot$ \citep[e.g.,][]{1999ApJ...518..356P}, the results shown here for $0.08~M_\odot$ are sufficient to establish the self-consistency. Since the outer regions of the disks are dominated by advective cooling (Figure \ref{fig2}(b)), they tend to maintain $Y_{\rm e} \sim 0.5$. By contrast, in Figure \ref{fig5}(c), even when the disk mass reaches $2~M_\odot$, the electron fraction stays low of $\sim 0.05$ – $0.2$ at $\dot m=5$, which represents the disk is neutron-rich. This behavior is consistent with an expanded inner region. As $\dot m$ increases, the neutrino-trapping radius $R_{\rm trap}$ and the neutrino-opaque radius move outward, electron degeneracy strengthens compared to the low-$\dot{m} $ case, and Urca processes favor electron capture, producing more neutrons and thus lowering $Y_{\rm e}$ \citep[e.g.,][]{2007ApJ...657..383C}. 
Moreover, \citet{2012ApJ...760...63L} estimated that a disk mass of order $\sim0.8~M_{\odot}$ is required to reproduce the extended emission observed in SGRBs for an initial BH of $m=3$, dimensionless spin $a_\ast=0.9$, and viscosity parameter $\alpha=0.01$; we therefore regard $0.8~M_\odot$ as an approximate upper bound for merger-formed disks under similar conditions. In Figure \ref{fig5}(b), For $\dot m=1$, a lighter disk ($<0.8~M_\odot$) has $Y_{\rm e}\lesssim0.3$, while the heavier disk shows gradual increase in $Y_{\rm e}$ with increasing accreted mass. 

Figure \ref{fig6} shows the radial distribution of $Y_{\rm e}$ under outflows of different strength. In Figure \ref{fig6}(a), with an outer boundary accretion rates $\dot{m}_{\rm outer}=0.1$ and corresponding accreted mass $0.05~M_\odot$, the region within $\sim200\ R_{\rm g}$ remains proton rich, similar to the no-outflow cases. Consistent with the logic applied to Figure \ref{fig5}(a), the convergence of $Y_{\rm e}\sim 0.5$ in the outer regions is a robust feature of NDAFs at low accretion rates. The presence of outflows does not alter the advection-dominated feature of the outer disks (Figures \ref{fig3}(b), (d) and (f)), ensuring that the self-consistency between NDAFs and collapsar progenitors remains valid as the disk mass scales up to $1~M_\odot$. For an outflow index $p=0.8$, the $Y_{\rm e}$ of inner-disk is systematically higher than for $p=0.5$. In Figure \ref{fig6}(b), with $\dot{m}_{\rm outer}=1$ and corresponding accreted mass $0.8~M_\odot$, the flow is overall neutron rich. For $a_*=0.1$ and $0.5$ with $p=0.8$, the combination of modest $\dot{m}_{\rm outer}$ and low spins implies that reaching $0.8~M_\odot$ occurs at larger radius, which can yield a higher $Y_{\rm e}\approx 0.38$, but this does not alter the global low-$Y_{\rm e}$ character. At smaller $\dot{m}_{\rm outer}$, the inner disk becomes cooler and less dense and thus lies in the neutrino-transparent regime. Under weak degeneracy ($\eta _{\mathrm{e}}<1$), the positron number density increases, and $\beta$-equilibrium shifts toward positron capture (i.e., Equation (\ref{eq40})), which raises the proton number density and hence increases the electron fraction, $Y_{\rm e} \equiv n_{\rm p}/(n_{\rm p}+n_{\rm n})$. Consequently, for the cases of $\dot{m}_{\rm outer}=0.1$ with strong outflows can accompany higher $Y_{\rm e} \sim 0.7$ \citep[e.g.,][]{2013ApJS..207...23X}, but the feature is not seen at $\dot{m}_{\rm outer}=1$.

Variations of BH spins have only a minor impact on these overall trends. Since $Y_{\rm e}$ is set by the local physical states of disks and a larger viscosity parameter ($\alpha \gtrsim 0.1$) could suppress disk neutronization in the high-spin case ($a_*=0.95$) \citep[e.g.,][]{2026ApJ...996..142H}, the global radial profiles show only weak spin dependence. We think that the mass accretion rate is a key factor shaping the disk properties. 

However, variations of the viscosity parameter $\alpha$ can significantly modify the disk structures. \citet{2026ApJ...996..142H} also showed that neutronization in the inner accretion flow becomes increasingly suppressed as $\alpha$ increases. The $\alpha$ can modify the critical ignition accretion rate, $\dot{M}_{\rm ign}$ \citep[e.g.,][]{2007ApJ...657..383C}, thereby affecting the radial distribution of disk physical properties. Studies of NDAF central engines typically adopt the viscosity parameter $\alpha = 0.01$ or $0.1$ \citep[e.g.,][]{2011MNRAS.410.2302Z,2010A&A...516A..16L}. In this work we fix $\alpha=0.1$, which facilitates comparison with a large fraction of the NDAF papers. Physically, varying $\alpha$ changes the radial transport and thus the density and thermal structure, pushing the disk into different pressure- and cooling-dominated regimes.  

\subsection{GRB Progenitors}

Based on the maps of the electron fraction distribution with accretion rates, spins, and outflow strength, we can establish a self-consistent correspondence between progenitors and electron fraction. Motivated by GRBs engine timescales, a simple estimation $t\sim M_{\rm disk}/\dot{M}$ suggests that longer-lived engines can be sustained by larger disk mass or lower accretion rates. Actually, the characteristic quantities of various evolutionary scenarios as shown in Table 1 of \citet{1999ApJ...518..356P} by reviewing most of theoretical and simulation studies on the different progenitors, including compact object mergers and massive collapsars. One can consider that the typical accretion rate of NDAFs without outflows arisen from mergers, $\sim 1~M_\odot~\rm s^{-1}$, is much larger than that from collapse, $\sim 0.1~M_\odot~\rm s^{-1}$; And the accreted mass is exactly the opposite. 

For the cases with lower accretion rates, the outcome is LGRBs with high $Y_{\rm e}$ (Figures \ref{fig5}(a) and \ref{fig6}(a)), which is a characteristic feature of accretion disks formed during the collapse of massive stars to BHs at the end of their evolution. Core collapses of massive stars is widely regarded as the progenitors of LGRBs \citep[e.g.,][]{2006ARAA..44..507W}, that has been confirmed observationally \citep[e.g.,][]{2003Natur.423..847H}.  

In recent studies, several works have suggested that collapsar disks can also yield low electron fractions. \citet{2025ApJ...985L..26I} found that magnetically arrested disk (MAD) could produce significant neutron-rich ($Y_{\rm e}<0.25$) ejecta at accretion rates larger than few solar mass per second, but collapsar progenitors with typical accretion rates, $\dot{m}\sim 0.1-1$, do not generate such ejecta.  \citet{2022ApJ...941..100S} argued that the inner regions of collapsar disks could become mildly electron degenerate, enabling self-neutronization via electron captures. \citet{2025arXiv250315729A} used GRMHD simulations to show that the collapse of very massive stars ($\gtrsim 10^2-10^4~M_{\odot }$) could produce higher accretion rates than ordinary collapsars. \citet{2019Natur.569..241S} performed three sets of simulations with different mass accretion rates and initial torus masses, and found that the disk midplane could reach ${Y_{\rm e}}\lesssim 0.2$ for accretion rates of order $\sim10^{-2}$ and $\sim10^{-3}$. In our calculations, for the cases with $\dot{m}_{\rm outer}=1$ and outflow $p=0.8$ (Figure~\ref{fig6}(b)), the inner accretion rates drop to the $\sim10^{-3}$, and the electron fractions likewise fall below $0.2$. These are not in contradiction with our finding of $Y_{\rm e}\sim 0.5$ at the outer boundary of the disk at low accretion rates as shown in Figure \ref{fig6}(a). While radial $Y_{\rm e}$ gradient exists across the disk due to local physical processes, the properties in the outer regions, where materials are initially supplied and outflows are likely launched \citep[e.g.,][]{2013MNRAS.435..502F}, remain inherently consistent with the progenitors. 

By contrast, the short-lived central engine generates a smaller disk and high instantaneous accretion rate. At higher accretion rates, the system produces SGRBs and exhibits low $Y_{\rm e}$ (Figures \ref{fig5}(c) and Figure \ref{fig6}(b)), which is a generic property of hyperaccreting disks formed after compact-object mergers \citep[e.g.,][]{2009MNRAS.396..304M}. Multiple lines of evidence, both observational \citep[e.g.,][]{2014ApJ...780..118F,2017ApJ...848L..12A} and theoretical \citep[e.g.,][]{1989Natur.340..126E,1992ApJ...395L..83N}, support an association between NS–NS/BH–NS mergers and SGRBs \citep[e.g.,][]{2020LRR....23....1M}. Consequently, low $Y_{\rm e}$ of NDAFs disks shows that SGRBs are associated with compact-object merger progenitors, whereas high $Y_{\rm e}$ indicates LGRBs are linked to the massive collapsar progenitors.

The disk outflows from NDAFs with different electron fraction naturally produce distinct electromagnetic counterparts, accounting for a subset of observed core-collapse supernovae (CCSNe) and kilonovae. In collapsar disks, neutrino irradiation and viscous driving typically yield higher $Y_{\rm e}$ (Figure \ref{fig6}(a)), enabling efficient synthesis of nuclei, especially ${}^{56}\mathrm{Ni}$, and powering CCSN emissions. \citet{2023ApJ...956..100F} showed that relativistic simulations formed in massive-star collapse demonstrate ${\gtrsim}0.1\,M_\odot$ of ${}^{56}\mathrm{Ni}$ and ejecta with floor values $Y_{\rm e}\!\gtrsim\!0.4$, consistent with Type~Ib/c supernova (SN) light curves. \citet{2003Natur.423..847H} reported that the spectrum of GRB 030329 confirmed the association between the LGRB and SN Ic, and provided further support for the hypothesis that LGRBs originated from the collapse of massive stars. By contrast, compact-object mergers disks generally eject neutron-rich materials ($Y_{\rm e}\!\lesssim\!0.2$, Figure \ref{fig6}(b)) that robustly drive heavy $r$-process nucleosynthesis and produce the lanthanide-rich, high-opacity red kilonova component ($\kappa\!\sim\!5$–$10\,\mathrm{cm^2\,g^{-1}}$). Modestly higher-$Y_{\rm e}$ outflows ($Y_{\rm e}\!\gtrsim\!0.25$) yield a lanthanide-poor blue component with much lower opacity \citep[$\kappa\!\sim\!0.5\,\mathrm{cm^2\,g^{-1}}$, e.g.,][]{2020MNRAS.496.1369T}. Observationally, AT2017gfo exhibits precisely such a two-component kilonova \citep[e.g.,][]{2017ApJ...851L..21V,2017Natur.551...64A}. In this work, we adopt the simplifying assumption that the electron fraction of disk-driven outflows and ejecta equals the local disk value at the launching layer. Neutrino absorption, and the competition between weak equilibration and viscous time-scales, can modify $Y_{\rm e}$ in outflows \citep[e.g.,][]{2013MNRAS.435..502F, 2015PhRvD..91l4021F}. Both outflow acceleration and entropy can also have an impact on the electron fraction \citep[e.g.,][]{2008ApJ...679L.117S}.

\section{Summary}\label{sec:Summary}

In this paper, we revisit the NDAF framework to investigate a self-consistent correspondence between the disk electron fraction and the GRB progenitors. Using a model that includes general relativity in the Kerr metric, neutrino microphysics, and nuclear statistical equilibrium, we considered accretion rates, BH spins, and outflow strength to compute disk structures and the distribution of $Y_{\rm e}$, and we checked these results against the characteristic electron fraction supplied by the progenitors. Our main findings are: (1) at low accretion rates, cooling in the outer disk is dominated by advection and neutrino cooling is relatively weak, so $Y_{\rm e}$ naturally remains near the symmetric value at the outer boundary of the disk, consistent with collapsar progenitors, and (2) at high accretion rates, the inner disk becomes efficiently neutrino cooled with mild electron degeneracy, which drives $Y_{\rm e}$ to low values at the outer boundary of the disk, consistent with compact-object merger progenitors. Variations in spins and outflow strength do not change these conclusions.

Regardless of its origin, the accretion rate of the BH hyperaccretion disk must undergo drastic variations. Initially, the accretion rate can be high enough to trigger NDAFs. Subsequently, it may drop sharply, leading to a regime where the Blandford-Znajek mechanism \citep[e.g.,][]{1977MNRAS.179..433B} replaces neutrino annihilation as the dominant power source for GRBs. Naturally, such a violent evolution of the accretion rate implies that baryons in the inner region of the disk transition from a free state to a bound state via Urca processes.

\section*{acknowledgments}
We thank Bao-Quan Huang, Jing-Tong Xing, Jiao-Zhen She, Yan-Qing Qi and Xiao-Yan Li for the helpful discussion. This work was supported by the National Natural Science Foundation of China (Grant Nos. 12494572, 12221003, and 12173031) and the Fund of National Key Laboratory of Plasma Physics (Grant No. 6142A04240201).

\end{document}